\documentclass[aps,pre,reprint,showpages,amsmath,preprintnumbers,nofootinbib]{revtex4-1}
\usepackage{graphicx}
\usepackage{subfigure}
\usepackage{comment}
\usepackage{url}
\usepackage{amsmath}
\usepackage{color}
\usepackage{here}
\begin{document}
\preprint{
KEK-TH-2164 
}
\date{\today}
\title{{Revisiting weak measurement in light of thermodynamics }}

\author{Daiki Ueda}
\affiliation{
KEK Theory Center, IPNS, Ibaraki 305-0801, Japan
}
\affiliation{
The Graduate University of Advanced Studies (Sokendai), Tsukuba, Ibaraki 305-0801, Japan.}

\author{Hikaru Ohta}
\affiliation{
KEK Theory Center, IPNS, Ibaraki 305-0801, Japan
}
\affiliation{
The Graduate University of Advanced Studies (Sokendai), Tsukuba, Ibaraki 305-0801, Japan.}

\author{Sumito Yokoo}
\affiliation{
KEK Theory Center, IPNS, Ibaraki 305-0801, Japan
}
\affiliation{
The Graduate University of Advanced Studies (Sokendai), Tsukuba, Ibaraki 305-0801, Japan.}

%%%%%%%%%%%%%%%%%%
\begin{abstract}
We investigate a weak measurement described by a von Neumann type interaction $\hat{A}\otimes \hat{p}^2$, where $\hat{A}$ is a system observable and $\hat{p}^2$ is a measurement pointer observable.  
%We propose a Maxwell's demon who appears in a weak measurement described by a von Neumann type interaction $\hat{A}\otimes \hat{p}^2$, where $\hat{A}$ is a system observable and $\hat{p}^2$ is a measurement pointer observable.  
%
We consider the weak measurement in terms of thermodynamics by adopting a mixed Gaussian state as a quantum state of the measurement pointer. 
We show that Maxwell's demon appears as a measure who carries out post-selections. 
It is found that, even if the demon only knows the weak value, a difference in the von Neumann entropy between the initial and final system can be the QC mutual information contents, which is the maximum amount of obtainable information by a measurement. 
Besides, our study indicates that a temperature of the system described by this interaction is amplified by weak value amplification.
In addition, we show that this demon can be realized in an atomic system. 

\end{abstract}
\pacs{
05.70.Ln
}

\maketitle

%Introduction
\section{Introduction}
The weak measurement with post-selection was proposed as an indirect measurement extracting an interesting physical quantity so called weak value\cite{PhysRev.134.B1410, Aharonov:1988ho}.
One of the interesting features of the weak value is that it can be amplified by a suitable post-selection.
Because of the large weak value, measurement pointer observables also can be amplified.  
This phenomenon is called the weak value amplification, which is applied to some precise measurements\cite{Ritchie:1991, Pryde:2005, Hosten787, PhysRevLett.102.173601} . 
The mechanism of the weak value amplification is roughly understood as extracting statistical outliers with large values of the measurement pointer observable by the post-selection.
The other is that the weak value can be obtained even if a measurement interaction of the indirect measurement is weak. 
Therefore, the weak value is the gained intermediate quantity of a system to be measured without disturbing the system.   
Then, the post-selection is an essential operation in all these properties of the weak value.

%actively discussed and well understood in terms of the indirect measurement.
The post-selection also plays an important role in feedback control.
The feedback control is a mechanical operation which is done after the post-selection depending on the outcome of the measurement.
The importance of the feedback control has been actively discussed in the context of the thermodynamics as Maxwell's demon\cite{Evans:1993, Gallavotti:1995, Jarzynski:1997, Crooks:1999, Jarzynski:2000, Hatano:2001, Seifert:2005, Sagawa:2008se, Sagawa:2010ge, Toyabe:2010ex, Seifert:2012}.
Maxwell's demon was proposed as a {\it gedankenexperiment} in the 19th century\cite{Maxwell:1871}.
{Many studies\cite{Szilard:1929, Landauer:1961, Bennett:1982, Piechocinska:2000, Lloyd:1997, Nielsen:1998, Maruyama:2005, Maruyama:2009} have indicated that information contents and a feedback controller play an essential role in order to understand the decrease in entropy of the thermodynamic system.}
{In particular, in \cite{Sagawa:2008se,Sagawa:2010ge}, the second law of thermodynamics including the feedback controller was formulated.}
Then, it would be interesting to consider the role of the weak value in Maxwell's demon.

{Stimulated by the given this situation, we reconsider the weak measurement in terms of the thermodynamics.}
For that purpose, we investigate the weak measurement with a von Neumann type interaction $\hat{A}\otimes \hat{p}^2$, where $\hat{A}$ is a system observable and $\hat{p}$ is a measurement pointer momentum.
Besides, we consider a mixed Gaussian state as a quantum state of the measurement pointer.
This setup makes a correspondence between the weak measurement with the post-selection and the thermodynamics clearer. 
From this point of view, the role of the weak value in Maxwell's demon of this setup also becomes clear.    
Then, it is shown that a difference in the von Neumann entropy becomes the QC mutual information contents even if the demon only knows the weak value of the observable $\hat{A}$. 
This result indicates that the maximum amount of obtainable information by the measurement are extracted in the weak limit where the quantum system is not disturbed.
Besides, we show that this setup may be realized by using an atomic system.
%.

 %
\section{measurement interaction}
We consider the weak measurement consisting of two quantum systems, $S$ and $M$ represented by the Hilbert spaces $\mathcal{H}_S$ and $\mathcal{H}_M$.     
From the point of view of quantum measurement, let $S$ and $M$ be a system to be measured and a measuring apparatus.  
In this paper, we focus on the following measurement interaction:
\begin{align}
\hat{H}= g \hat{A}\otimes \hat{p}_L^2,\label{eq:int}
\end{align} 
where $\hat{A}$ is an observable in the system $S$, $\hat{p}_L$ is a momentum of the system $M$ defined in a one-dimension with finite range $L$ and $g$ is a coupling parameter.
The operator of the momentum in the system $M$ is expressed as $\hat{p}_L \equiv \sum_i p_{i,L}{|p_{i,L}\rangle}{\langle p_{i, L}|}$, where ${|p_{i,L}\rangle}$ is $i\text{--}$th eigenstate and $p_{i,L}$ is its eigenvalue.
Note that the measurement pointer observable of the measurement interaction is not $\hat{p}_L$ but $\hat{p}_L^2$.
As discussed in the next section, the measurement interaction Eq.~\eqref{eq:int} enables us to understand the weak measurement with the post-selection in terms of the thermodynamics.

Before proceeding further, we explain that the interaction Eq.~\eqref{eq:int} is realized by using an atomic system.
Let us consider a motion of an atom with binding energy.
Then, the entire Hilbert space consists of two quantum systems $\mathcal{H}_S$ and $\mathcal{H}_M$, where $S$ is a system consisting of the atomic energy levels and $M$ corresponds to the atomic center-of-mass position system.     
%
%The Hilbert space of this system is expressed as $\mathcal{H}=\mathcal{H}_E\otimes \mathcal{H}_P$ where $\mathcal{H}_E$($\mathcal{H}_P$) is the Hilbert space of atom's energy levels(atom's center-of-mass position)\footnote{In view of the quantum measurement, $\mathcal{H}_E$ corresponds to a measured system, and $\mathcal{H}_P$ corresponds to a probe system.}.
%
For simplicity, we consider an atom in a one-dimensional space\footnote{ An extension to a three-dimensional space is straightforward.}.
The Hamiltonian of this atomic system is given as
\begin{align}
\hat{H}=\hat{H}_{\rm atom} + \hat{H}_K(L)+\frac{\Delta m}{4\bar{m}^2} \hat{A} \otimes \hat{p}_L^2,\label{eq: ham}
\end{align}
where $\hat{H}_{\rm atom}$ is the Hamiltonian of an atom in a rest frame, $\hat{H}_K(L)\equiv\hat{p}^2_L/2\bar{m}$ is the Hamiltonian corresponding to a kinetic motion of the atomic center-of-mass, $\bar{m}$ is an averaged mass of the atom including its binding energy, and $\hat{p}_L$ is a momentum of the atomic center-of-mass in a one-dimension with finite range $L$.   
The last term in Eq.(\ref{eq: ham}) comes from the mass difference between the atomic ground and excited states.
The detail definition is given later.
Here, we define the Hamiltonian of an atom: $\hat{H}_{\rm atom}= \sum_{n=1}^{\infty} E_n {|n \rangle}{\langle n|}$, where $E_n$ is the $n\text{--}$th atomic energy eigenvalue, and ${| n\rangle}$ is the $n\text{--}$th atomic energy eigenstate.
For example, ${|1\rangle}$ and ${|2\rangle}$ correspond to the ground and the first excited state of the atom respectively, and their difference comes from the hyperfine splitting.
Hereafter we focus on the Hilbert space $\mathcal{H}_S$ spanned by ${| 1\rangle}$ and ${| 2\rangle}$. 

Finally, we discuss an origin of the last term in Eq.(\ref{eq: ham}).
The binding energy of the atom contributes to the kinetic energy of the atomic center-of-mass because of the mass-energy equivalence\cite{Rosi:2017qu,10.1093/ptep/ptz019}.
The effect of the binding energy is included as a mass operator for the atom, which is expressed as $\hat{m}=\bar{m} -\Delta m \hat{A}/2$, where $\bar{m}\equiv (m_1+m_2)/2$, $\hat{A}\equiv\left({|1\rangle}{\langle 1|} -{|2 \rangle}{\langle 2|} \right)$ and $\Delta m\equiv m_2-m_1$.
Thus, the atomic kinetic energy is approximated as $\hat{p}^2/2\hat{m}\simeq \hat{H}_K +\Delta m \hat{A} \otimes \hat{p}^2/4 \bar{m}^2$ by expanding the mass operator with respect to the mass difference $\Delta m$.
%
%Even if in a case that many-body system consisting the cold atoms described by Eq.~(\ref{eq: ham}), for simplicity, we assume that interactions between the atoms are negligible.
%

\section{Weak measurement with post-selection}
We consider the weak measurement with the measurement interaction Eq.~\eqref{eq:int} in terms of the thermodynamics.
For that purpose, we adopt the following mixed Gaussian state as the initial quantum state of the measurement pointer $M$:
\begin{align}
\hat{\rho}^M_i ={\rm exp}\left(-d^2 \hat{p}^2_L/2 \right)/Z_{d,L},\label{eq:ins}
%\sum_i e^{-d^2 p_{i,L}^2/2} {|p_{i,L}\rangle}{\langle p_{i,L}|}/Z_{d,L}
\end{align}
where $Z_{d,L} \equiv {\rm Tr}_M \left({\rm exp}\left(-d^2 \hat{p}^2_L/2 \right) \right)$, and $d^{-1}$represents a width of the momentum of the measurement pointer.
Here, ${\rm Tr}_M$ stands for a trace with respect to $\mathcal{H}_M$

Let us consider the weak measurement by using the measurement interaction Eq.~\eqref{eq:int} and the initial quantum state of the measurement pointer Eq.~\eqref{eq:ins}.
In the weak measurement with the post-selection, the time evolution of the entire system is divided into four processes.
Each process is the followings: 

\begin{enumerate}

\item {\it Pre-selection.}
At time $t_i$, the entire quantum state is prepared in  
\begin{align}
\hat{\rho}_1\equiv \hat{\rho}^S_{i}\otimes \hat{\rho}^M_i,
\end{align}
where $\hat{\rho}^S_{i}$ is the pre-selected state in the system $S$ and $\hat{\rho}^M_i$ is the initial state of the system $M$ defined in Eq.~\eqref{eq:ins}.

\item {\it Unitary time evolution.}
From $t_i$ to $t_f$, the entire state evolutes by the unitary operator corresponding to the Hamiltonian Eq.(\ref{eq:int}) as 
\footnotesize
\begin{align}
\hat{\rho}_2 &\equiv e^{-iT\hat{H}}\hat{\rho}_i  e^{+iT\hat{H}}\notag
\\
&\tiny{\simeq \left(1-ig T \hat{A}\otimes \hat{p}_L^2 \right) \hat{\rho}_i \left(1+i g T \hat{A}\otimes \hat{p}_L^2 \right)+\mathcal{O}( g^2)},\label{eq: timeev}
\end{align}
\normalsize
where we defined $T\equiv t_f-t_i$.
Here and hereafter, we put $\hbar=1$ for brevity.
In order to investigate the weak measurement with the post-selection, the time evolution operator is expanded up to the first order of $g$.

\item {\it Measurement on $S$.}
At time $t_f$, the measurement on the system $S$ is performed.
We adopt a projection-valued measure (PVM) measurement described by a projection operator, $\hat{D}_k\equiv {|k\rangle}{\langle k|}\otimes \hat{1}_M$, where the state ${|k\rangle}\in \mathcal{H}_S$ and $\hat{1}_M$ is the identity operator on $\mathcal{H}_M$.   
After the measurement, the state of the entire system is obtained as
\begin{align}
\hat{\rho}_3 \equiv\sum_k \sqrt{\hat{D}_k}\hat{\rho}_2 \sqrt{\hat{D}_k}=\sum_k p_k  \hat{\rho}^{(k)}(L),
\end{align} 
where
\begin{align}
p_k &\equiv {\rm Tr}(\hat{D}_k \hat{\rho}_2)\notag
\\
&={\langle k|}\hat{\rho}^S_{i}{|k\rangle} \left(1+g T\cdot{\rm Im}~A^W_{k} \right) +\mathcal{O}(g^2)\label{eq:pk},
\\
\hat{\rho}^{(k)}(L)&\equiv \sqrt{\hat{D}_k}\hat{\rho}_2\sqrt{\hat{D}_k}/p_k\notag
\\
&={|k\rangle}{\langle k|}\otimes \hat{\rho}^{(k)}(L)+\mathcal{O}(g^2),\label{eq: rhok}
\end{align}
Here, $A^W_{k} \equiv {\langle k|} \hat{A} \hat{\rho}^S_{i}{|k\rangle}/{{\langle k|} \hat{\rho}^S_{i}{|k\rangle}}$ is the weak value of $\hat{A}$, and $\hat{\rho}^{(k)}(L)$ is the state of $M$ after the system $S$ is observed as $k$:
 \begin{align}
 \hat{\rho}^{(k)}(L)={\rm exp}\left(-d_k^2 \hat{p}_L^2/2 \right)/Z_{d_k,L},
 %\sum_i e^{-d^2_{k}\cdot p_{i ,L}^2/2} {|p_{i,L} \rangle}{\langle p_{i ,L}|}/Z_{d_k,L},
 \end{align}
where $Z_{d_k,L}\equiv {\rm Tr}_M \left({\rm exp}\left(-d_k^2 \hat{p}_L^2/2 \right)\right)$, and $d^{2}_k \equiv d^2\cdot \left(1-4 g T {\rm Im}~A^W_k /d^2 \right)$. 
It is interesting that the weak value $A^W_k$ appears in $d_k$.

\item {\it Post-selection.}
After the measurement on the system $S$, one of the measured states is selected.
For example, let $k\in\{\psi,\phi\}$ be the measured outcomes and suppose that the post-selected state is ${|\phi\rangle}$.
Then, the quantum state of $M$ after the post-selection is obtained as
\begin{align}
\hat{\rho}^{(\phi)}(L) ={\rm exp}\left(-d_{\phi}^2 \hat{p}_L^2/2 \right)/Z_{d_{\phi},L}
%\sum_i e^{-d^2_{\phi}\cdot p_{i,L}^2} |p_{i,L}\rangle {\langle p_{i ,L}|}/Z_{d_{\phi},L}.
\end{align}

\end{enumerate}

After the post-selection, a variance of the momentum of $M$ is obtained as
\begin{align}
{\rm Tr}_M \left(\hat{p}^2 \hat{\rho}^{(\phi)}(L) \right) - \left({\rm Tr}_M \left(\hat{p} \hat{\rho}^{(\phi)}(L) \right)\right)^2 =d_{\phi}^{-2}.\label{eq:var1}
\end{align}
The right hand side of Eq.~\eqref{eq:var1} includes the imaginary part of the weak value $A^W_{\phi}$.
Therefore, the variance of the momentum of $M$ can be amplified by the weak value amplification.

Next, we move to the thermodynamic understanding of the above weak measurement.
Let $S$ and $M$ be the quantum system corresponding to a thermometer and a thermodynamic system to be measured.
Note that the roles of the quantum systems are swapped compared with the conventional indirect measurement.  
In addition, let us suppose that the width of the initial state of $M$ is a temperature of $M$.
Then, the weak value amplification of the variance of $M$ represents that the post-selected state $|\phi\rangle$ corresponds to a state of the thermometer which reads a high temperature region.
If the weak value amplification occurs by the post-selected state $|\phi\rangle$, the variance of the momentum of $M$ is not amplified by selecting ${|\psi\rangle}$ as the post-selected state because of a relational expression between ${\rm Im}~A^W_{\phi}$ and ${\rm Im}~A^W_{\psi}$: 
\begin{align}
{\rm Im}~A^W_{\psi}= -{\rm Im}~A^W_{\phi} \frac{{\langle \phi|\hat{\rho}^S_i|\phi\rangle}}{{\langle \psi|\hat{\rho}^S_i|\psi\rangle}},\label{eq:rel1}
\end{align}  
Therefore, the post-selected state $|\psi\rangle$ corresponds to a low temperature of the thermometer.
It is remarkable that measuring the temperature of $M$ corresponds to measuring the weak value. 
These results show that the thermodynamic system consisting of particles can be divided into two parts, high and low temperature systems depending on the weak value.
This mechanism is reminiscent of Maxwell's demon.
In the next section, we discuss a role of the weak value in the demon of this quantum system by restricting to $k\in\{\psi,\phi\}$.

%
%%%%%%%%%%%%%%%%%%%%%%%%%%%%%%%%%%%%%%%%%%%%
%\begin{figure}[t]
%\begin{center}
%\includegraphics[scale=0.36]{Fig.eps}
%\caption{
%The difference in the von Neumann entropy (upper) and the imaginary part of the weak value ${\rm Im} \Delta m^W_{\psi,\phi,\Delta t}$ (middle and lower) as a function of $\delta$.
%
%Red, blue and Green curves correspond to $\epsilon=10^{-3}, 10^{-1}$ and $2\times 10^{-1}$ respectively. 
%}
%\label{fig:plot1}
%\end{center}
%\end{figure}
%%%%%%%%%%%%%%%%%%%%%%%%%%%%%%%%%%%%%%%%%%%%

%The processes 3 and 4 can be performed by the Stern-Gerlach experiment with an external electric field, and paths of the atom are split depending on the atom's electric dipole moment. 
%
%In the context of the weak value amplification, focusing only on the specific path is often regarded as the post-selection, which include the implicit mechanical operations depending on the outcome of the measurement.

\section{Maxwell's demon}
We construct a setup where Maxwell's demon appears by the weak measurement discussed in the previous section.
In such a setup, feedback controls are essential in the time evolution of the entire system.
Here, we replace the stage 4, {\it Post-selection} in the previous section with the feedback control :
\begin{enumerate}
\setcounter{enumi}{3}
\item {\it Feedback control.}
After the measurement on $S$, the entire state evolves by mechanical operations depending on the observed state $k\in\{\psi,\phi\}$.
After the feedback control, we obtain the state of the entire system:
\begin{align}
\hat{\rho}_4 \equiv \sum _{k\in\{\psi,\phi\}} p_k \hat{\rho}_4^{(k)}.
\end{align}
We adopt a mechanical operation on $M$: 
\begin{align}
& \hat{\rho}^{(k)} (L)\to \hat{\rho}_4^{(k)} =  \hat{U}'_k\hat{\rho}^{(k)} (L_{k})\hat{U'}_k^{\dagger},
\end{align}
where 
\begin{align}
&L_{\psi}= L\cdot {\rm Im}~ A^W_{\phi} /({\rm Im} A^W_{\phi}-{\rm Im} A^W_{\psi}),\label{eq: psil}
\\
&L_{\phi}= L\cdot {\rm Im}~ A^W_{\psi} /({\rm Im} A^W_{\psi}-{\rm Im} A^W_{\phi}),\label{eq: phil}
\end{align}
and $\hat{U'}_k\equiv e^{-ig_k' \hat{x}}$.
This operation corresponds to the isothermal process.
For convenience, we inserted the unitary operator $\hat{U'}_k$ corresponding to the division of the thermodynamic system $M$ into two parts, however, this insertion does not affect the difference in the von Neumann entropy.
It is noteworthy that the weak value appears in $L_k$ and $L_{\psi}+L_{\phi}=L$ is satisfied.
This means that a measure who carries out the feedback control knows the weak value.  
%
%The detail is discussed latter.
%
%
Then, we have the following state:
\begin{align}
\hat{\rho}_4 = \sum_{k\in \{\psi,\phi\}} p_k\hat{U}'_k \hat{\rho}^{(k)}(L_k)\hat{U'}_k^{\dagger},
\end{align}
where
\begin{align}
 \hat{\rho}^{(k)}(L_k)= {\rm exp}\left(-d_{k}^2 \hat{p}_{L_k}^2/2 \right)/Z_{d_{k},L_k}
 %\sum_i e^{-d_k^2 \cdot p_{i,L_k}^2} {|p_{i,L_k} +g_k\rangle} {\langle p_{i,L_k}+g_k|}/Z_{d_k,L_k},
\end{align}
and $p_k$ was defined in Eq.~\eqref{eq:pk}.
This feedback control represents a change of a volume of $M$ depending on the observed state $k$.
An important point is that the feedback control is done by selecting one of the measured states.
Therefore, the feedback control includes the post-selection if we focus on one of the mechanical operations. 
This fact connects the weak value to the demon.

\end{enumerate}
%
%Fig.~\ref{} schematically express the time evolution of the entire system.
%
In this setup, the demon appears as a measure who carries out post-selection by using the weak measurement with the measurement interaction Eq.~\eqref{eq:int}.  
This setup is similar to the Szilard engine\cite{Szilard:1929}, where the difference in the von Neumann entropy between the initial and final system $\Delta S = S(\hat{\rho}_4)- S(\hat{\rho_1})$ becomes the QC mutual information content defined as
\begin{align}
I (\hat{\rho}_2,\{k \})&\equiv S(\hat \rho_2)-\sum_k p_k\ln p_k \notag \\
&+\sum_k{\rm Tr}\left[\sqrt{\hat D_k}\hat \rho_2\sqrt{\hat D_k}\ln \left(\sqrt{\hat D_k}\hat \rho_2\sqrt{\hat D_k}\right)\right].
\end{align}
Here, the von Neumann entropy of $\hat \rho_2$ is defined as $S(\hat \rho_2) \equiv -{\rm Tr}\left(\hat{\rho}_2\ln\hat{\rho}_2 \right)$.
$I (\hat{\rho}_2,\{k \})$ is the maximum amount of information that a measure can obtain by the measurement and gives the lower bound for $\Delta S$ in the general feedback control.
Actually, the difference in the von Neumann entropy in our system is obtained as 
\begin{align}
\Delta S \equiv S(\hat{\rho}_4)- S(\hat{\rho_1})=-S(\hat{\rho}^S_i)  =-I (\hat{\rho}_2,\{k \})\label{eq:difS}.
\end{align}
See Appendix for the detail calculations.
%
%The right hand side of Eq.~\eqref{eq:difS} is the QC mutual information contents, which is the information about the measured entire system that has been obtained by measurement in the stage $3$.
%
It is interesting that the difference in the von Neumann entropy can become the QC mutual information contents by the feedback control on $M$ depending on the weak value.
This implies that the maximum amount of obtainable information can be extracted by the feedback control on $M$, even in the weak limit where the quantum system is not disturbed.
It is noteworthy that the suitable feedback control on $M$\footnote{The suitable feedback control on $S$ would be clear, because the measurement on $S$ is not weak.} can be clear by measurement on $S$ in the weak limit.
%
%This result is one of merits of considering the weak measurement in terms of the thermodynamics. 

\section{{Atomic system}}
\label{sec:Res}
We consider Maxwell's demon by using the atomic system described by the Hamiltonian of Eq.~\eqref{eq: ham}.
In this system, the atomic energy levels and the atomic center-of-mass position correspond to the Hilbert space of the thermometer and the thermodynamic system, respectively.    
The time evolution of the entire system is divided into four processes:
\begin{enumerate}

\item {\it Pre-selection.}
At  time $t_i$, the entire quantum state is prepared in
\begin{align}
\hat{\rho}_1\equiv \hat{\rho}^S_{i,\delta}\otimes \hat{\rho}^M_i,
\end{align}
where $\hat{\rho}^S_{i,\delta}$ is the pre-selected state.
We consider the following initial state:
\begin{align}
&\hat{\rho}^S_{i,\delta}\equiv (1-\epsilon) {|\psi_{\delta}\rangle}{\langle \psi_{\delta}|} +\epsilon {|\phi_{\delta}\rangle}{\langle \phi_{\delta}|},\label{eq: Epr}
\\
&\hat{\rho}^M_i \equiv {\rm exp}\left(-\beta \hat{H}_K(L)\right)/Z_{\beta,L},\label{eq: EpM}
\end{align}
where $Z_{\beta,L}\equiv {\rm Tr}_M \left( {\rm exp}\left(-\beta \hat{H}_K(L)\right)\right)$, ${|\psi_{\delta}\rangle}\equiv(e^{+i\delta}{|1\rangle}+e^{-i\delta}{|2\rangle})/\sqrt{2}$ and ${|\phi_{\delta}\rangle}\equiv(e^{+i\delta}{|1\rangle}-e^{-i\delta}{|2\rangle})/\sqrt{2}$.
$\hat{\rho}^M_i$ is a mixed Gaussian state with a variance $\beta/2\bar{m}$.
The states $|\psi_{\delta=0}\rangle$ and $|\phi_{\delta=0}\rangle$ are the eigenstates of an atomic electric dipole moment, which can be prepared by applying an electric field. 
In Eq.~\eqref{eq: Epr}, the states $|\psi_{\delta}\rangle$ and $|\phi_{\delta}\rangle$ are mixed by a parameter $\epsilon$.
The parameter $\delta$ is introduced in order to control the weak value.
\item {\it Unitary time evolution.}
From $t_i$ to $t_f$, the time evolution of the entire state described by the Hamiltonian Eq.(\ref{eq: ham}) is obtained as 
\footnotesize
\begin{align}
\hat{\rho}_2 &\equiv e^{-iT\hat{H}}\hat{\rho}_1  e^{+iT\hat{H}}\notag
\\
&\tiny{\simeq \left(1-i\chi \hat{A}\otimes \beta\frac{\hat{p}^2}{2\bar{m}} \right)\hat{\rho}_{1}(T) \left(1+i\chi \hat{A}\otimes \beta\frac{\hat{p}^2}{2\bar{m}} \right)+\mathcal{O}( \chi^2)},\label{eq: timeev}
\end{align}
\normalsize
where $T\equiv t_f-t_i$, $\chi\equiv T \Delta m/2\bar{m}\beta$ and $\hat{\rho}_1(T)\equiv e^{+iT\Delta E \hat{A}/2} \hat{\rho}_1 e^{-iT\Delta E \hat{A}/2}$ are defined.
Here, the Hamiltonian of the atom is rewritten as $\hat{H}_{\rm atom}=(E_1+E_2)/2-\Delta E \hat{A}/2$, where $\Delta E\equiv E_2-E_1$.
Hereafter, focusing on a short time regime $\Delta t \cdot\Delta E/2 \ll \delta$, time evolution coming from $\hat{H}_{\rm atom}$ is neglected.
In Eq.(\ref{eq: timeev}), the time evolution operator is expanded up to the first order of $\chi$ in order to investigate the relation between the weak value and Maxwell's demon.

\item {\it Measurement on $S$.}
At the time $t_f$, the measurement on the system $S$ is performed.
We adopt a PVM measurement described by a projection operator $\hat{D}_k\equiv {|k\rangle}{\langle k|}\otimes \hat{1}_M$ for $k=\psi$ and $\phi$.
This PVM measurement represents the projection measurement of the atomic electric dipole moment.
After the measurement, the state of the entire system is given as
\begin{align}
\hat{\rho}_3 \equiv\sum_k \sqrt{\hat{D}_k}\hat{\rho}_2 \sqrt{\hat{D}_k}=\sum_k p_k  \hat{\rho}^{(k)}(L) .
\end{align} 
Here, we defined
\begin{align}
&p_k\equiv {\rm Tr}(\hat{D}_k \hat{\rho}_2)\notag
\\
&={\langle k|}\hat{\rho}^E_{i,\delta_T}{|k\rangle} \left(1+\chi\cdot{\rm Im}~A^W_{k} \right) +\mathcal{O}(\chi^2),
\\
&\hat{\rho}^{(k)}(L)\equiv\sqrt{\hat{D}_k}\hat{\rho}_2\sqrt{\hat{D}_k}/p_k\notag
\\
&={|k\rangle}{\langle k|}\otimes{\rm exp}\left(-\beta_{k} \hat{H}_K(L)\right)/Z_{k,L}+\mathcal{O}(\chi^2),\label{eq: rhok}
\end{align}
$\beta_{k} \equiv\beta\cdot \left(1-2\chi\cdot {\rm Im}~A^W_{k}\right)$ and $Z_{k,L}\equiv {\rm Tr}_M \left({\rm exp}\left(-\beta_{k} \hat{H}_ K(L)\right) \right)$. 
%
%Here, notice that the approximation with respect to the first order of $g$ is justified if a quantity $g\cdot {\rm Im}~A^W_k$ is smaller than unity. 
%
%Thus, in parameter regions where the approximation is valid, $\beta_k$ takes positive value. 
%
For convenience, we defined as $\delta_T\equiv \delta +T\cdot \Delta E/2$.
Then, the weak value of the atomic energy level was defined as
\begin{align}
A^W_{k}&\equiv \frac{{\langle k|} \hat{A} \hat{\rho}^S_{i,\delta_T}{|k\rangle}}{{\langle k|} \hat{\rho}^S_{i,\delta_T}{|k\rangle}},\label{eq: weak}
\end{align}
which appears in the temperature $\beta_k$.

\item {\it Feedback control.}
After the measurement on $S$, the feedback control depending on the atomic electric dipole moment is performed.
%
%We assume that $\hat{U}_k$ for $k=\psi,\phi$ is the corresponding mechanical operations described by a unitary operator.
%
After the feedback control, the state of the entire system is given as
\begin{align}
\hat{\rho}_4\equiv \sum _k  p_4\hat{\rho}_4^{(k)}.
\end{align}
Similar to the previous section, we adopt a mechanical operator:
 $\hat{\rho}_4^{(k)}= \hat{\rho}^{(k)}(L_k),$
where $L_{\psi,\phi}$ was defined in Eq.~\eqref{eq: psil} and \eqref{eq: phil}.
Since the mechanical operation $\hat{U}'_k$ does not affect the entropy, we discarded it. 

\end{enumerate}
Eq.(\ref{eq: rhok}) shows that thermodynamic equilibrium states at temperature $\beta_{k}$ can be prepared by focusing on the state $k$.
The temperature $\beta_k$ depends on the weak value, which can be amplified by the weak value amplification.
As shown in Appendix, the weak values are obtained as
\begin{align}
&A^W_{\psi}=i \frac{(1-2\epsilon)\sin\delta_T\cos\delta_T}{\epsilon \sin^2\delta_T +(1-\epsilon)\cos^2 \delta_T },~{\rm for}~k=\psi,
\\
&A^W_{\phi}=i \frac{-(1-2\epsilon)\sin\delta_T\cos\delta_T}{\epsilon \cos^2\delta_T +(1-\epsilon)\sin^2 \delta_T },~{\rm for}~k=\phi.
\end{align}
Because of Eq.~\eqref{eq:rel1}, the imaginary part of the weak value ${\rm Im}~A^W_{\psi}$ always takes the opposite sign of ${\rm Im}~A^W_{\phi}$.
In particular, in a region near $\epsilon\simeq 0$ and $\delta_T \simeq \pi/2\cdot n$ for $n=0,1,2,\cdots$, ${\rm Im}~A^W_{\phi}$ takes a large positive value by the weak value amplification, and ${\rm Im}~A^W_{\psi}$ takes a negative value near zero.
This means that many-body system consisting of the atoms can be divided into a high and low temperature region if the weak value amplification occurs.   
%
%Here, we emphasize that this temperature amplification occurs thanks to the interaction $\hat{A}\otimes \hat{p}^2$.   
%
%This phenomena does not occur by the interaction $\hat{A}\otimes \hat{p}$.    

The difference in the von Neumann entropy between the initial and final system is also obtained as $\Delta S\equiv S(\hat{\rho}_4)-S(\hat{\rho}_1)= -I (\hat{\rho}_2,\{k \})$. 
%
%\begin{widetext}
%\begin{align}
%\Delta S\equiv S(\hat{\rho}_f)-S(\hat{\rho}_i)&= -I_{QC},\label{eq: difent}
%H(\{{\langle k|\hat{\rho}^E_{i,\delta_T}|k\rangle }\})-g\cdot \frac{{\rm Im}~A^W_{\psi}{\rm Im}~A^W_{\phi}}{{\rm Im}~A^W_{\phi}-{\rm Im}~A^W_{\psi}} \ln \left[ - \frac{{\rm Im}~A^W_{\phi}}{{\rm Im}~A^W_{\psi}}\right] -S(\hat{\rho}^E_i)
%+\mathcal{O}(g^2)
%,\label{eq: difent}
%\\
%H(\{{\langle k|\hat{\rho}^E_{i,\delta_T}|k\rangle }\}) &=-\frac{{\rm Im}~A^W_{\psi}}{{\rm Im}~A^W_{\psi} -{\rm Im}~A^W_{\phi}} \ln\left[\frac{{\rm Im}~A^W_{\psi}}{{\rm Im}~A^W_{\psi} -{\rm Im}~A^W_{\phi}}\right] -\frac{{\rm Im}~A^W_{\phi}}{{\rm Im}~A^W_{\phi}-{\rm Im}~A^W_{\psi}}\ln\left[\frac{{\rm Im}~A^W_{\phi}}{{\rm Im}~A^W_{\phi}-{\rm Im}~A^W_{\psi}}\right],\label{eq: dien1}
%\end{align}
%\end{widetext}
%
%where $S(\hat{\rho})\equiv -{\rm Tr}(\hat{\rho}\ln\hat{\rho})$ is the von Neumann entropy and $H(\{p_k\})\equiv -\sum_{k
%\in \{ \psi,\phi\}}p_k \ln p_k $ is the Shannon information content.
%
See Appendix for the detail calculations.
%
%Eq.(\ref{eq: difent}) and Eq.(\ref{eq: dien1}) shows that, except for parameters in the Hamiltonian, the difference in the von Neumann entropy is controlled by the weak value. 
%
%In this sense, it is implied that Maxwell's demon controls the weak value in this system. 
%
%As mentioned in the next section, the difference of the von Neumann entropy can be negative, if the weak value amplification occurs.
%
%As discussed in \cite{Sagawa:2008se}, the difference in the von Neumann entropy can be negative by measurement processes, and that is bounded by the QC-mutual information content like $\Delta S \geq - I(\hat{\rho}_i,\{p_k\})$, which represents the information obtained by the measurement, and $I(\hat{\rho}_i,\{p_k\})$ satisfies $0 \leq I(\hat{\rho}_1, \{p_k\}) \leq H (\{p_k\})$.
%
In this setup, the QC-mutual information content of the system is obtained as $I (\hat{\rho}_2,\{k \}) =S(\hat{\rho}^E_i)+\mathcal{O}(\chi^2)$. 

\section{Conclusions}
In this paper, we have investigated the weak measurement described by the von Neumann type interaction $\hat{A}\otimes \hat{p}^2$, where $\hat{A}$ is an observable in a system to be measured $S$ and $\hat{p}$ is a momentum of a measurement pointer $M$.
By adopting a mixed Gaussian state as the initial quantum state of $M$, we have reconsidered the weak measurement in terms of the thermodynamics.
From the point of view of the thermodynamic understanding, we have shown that $S$ and $M$ can be regarded as a thermometer and a thermodynamic system respectively.
Then, we have seen that the temperature of $M$ can be amplified by the post-selection.

Besides, we have investigated a role of the weak value in Maxwell's demon by focusing on a fact that the feedback control involves the post-selection.  
Then, we have seen that the difference in the von Neumann entropy can become the QC mutual information contents even if the demon only knows the weak value of $\hat{A}$.
This result means that the demon can extract the maximum amount of obtainable information by the feedback control on $M$ even in the weak limit where the quantum system is not disturbed.
In addition, we proposed a system described by the von Neumann type interaction by using an atomic system.
Our study indicates that the weak value plays an important role in the extraction of the obtainable information contents by the feedback control including the post-selection. \\

\section*{Acknowledgement}
We are grateful to Kiyoharu Kawana for the collaboration in the early stage of this project.
We greatly appreciate many valuable conservations with our colleagues, Ryota Kojima, Naoto Kan, Takato Mori, Yuichiro Mori, Katsumasa Nakayama, Yoshinori Tomiyoshi.

%%%%%%%%%%%%%%%%%%%%%%%%%%%%%%%%%%%%%%%%%%%%%%%%%%%%%%%% 

\begin{widetext}
\appendix

\section{Calculation of weak value}
In this section, we give detailed calculations of the weak value $A^W_k$.
The denominator in Eq. (\ref{eq: weak}) is given as
\begin{align}
{\langle k|\hat{\rho}^S_{i,\delta_T} |k\rangle}=(1-\epsilon)\left|{\langle k|\psi_{\delta_T}\rangle} \right|^2 +\epsilon \left|{\langle k|\phi_{\delta_T}\rangle} \right|^2,~~~{\rm for}~k=\psi,\phi.\label{eq: den1}
\end{align}
For $k=\psi$ and $k=\phi$, Eq.(\ref{eq: den1}) is calculated as
\begin{align}
{\langle \psi|\hat{\rho}^S_{i,\delta_T} |\psi\rangle}=& (1-\epsilon) \cos^2 \delta_T +\epsilon \sin^2 \delta_T,\label{eq: den2}
\\
{\langle \phi|\hat{\rho}^S_{i,\delta_T} |\phi\rangle}=& (1-\epsilon) \sin^2 \delta_T +\epsilon \cos^2 \delta_T.\label{eq: den3}
\end{align}
Besides, the numerator in Eq. (\ref{eq: weak}) is given as
\begin{align}
{\langle k|\hat{A} \hat{\rho}^S_{i,\delta_T}|k \rangle}&=(1-\epsilon){\langle k|1\rangle} {\langle 1|\psi_{\delta_T}\rangle}{\langle \psi_{\delta_T}|k\rangle} 
+\epsilon {\langle k|1\rangle} {\langle 1|\phi_{\delta_T}\rangle} {\langle \phi_{\delta_T}|k\rangle}\notag
\\
&-(1-\epsilon) {\langle k|2\rangle} {\langle 2|\psi_{\delta_T}\rangle} {\langle \psi_{\delta_T}|k\rangle}
-\epsilon {\langle k|2\rangle} {\langle 2|\phi_{\delta_T}\rangle} {\langle \phi_{\delta_T}|k\rangle},~{\rm for}~k=\psi,\phi.\label{eq: nu1}
\end{align}
For $k=\psi$ and $k=\phi$, Eq.(\ref{eq: nu1}) is calculated as
\begin{align}
&{\langle \psi|\hat{A} \hat{\rho}^S_{i,\delta_T}|\psi\rangle}= i (1-2\epsilon) \sin\delta_T \cos \delta_T,\label{eq: nu2}
\\
&{\langle \phi|\hat{A} \hat{\rho}^S_{i,\delta_T}|\phi\rangle}=- i (1-2\epsilon) \sin\delta_T \cos \delta_T.\label{eq: nu3}
\end{align}
Combing Eq. (\ref{eq: den2}), Eq. (\ref{eq: den3}), Eq. (\ref{eq: nu2}) and Eq. (\ref{eq: nu3}), the weak values are given as
\begin{align}
&A^W_{\psi}=i\frac{(1-2\epsilon)\sin\delta_T \cos\delta_T}{(1-\epsilon)\cos^2\delta_T +\epsilon \sin^2 \delta_T},
\\
&A^W_{\phi}=-i\frac{(1-2\epsilon)\sin\delta_T \cos\delta_T}{(1-\epsilon)\sin^2\delta_T +\epsilon \cos^2 \delta_T}.
\end{align}

Here, the weak values satisfy a following relation:
\begin{align}
\sum_{k=\psi,\phi}  {\langle k| \hat{\rho}^S_{i,\delta_T}|k \rangle}\cdot {\rm Im}A^W_k=\sum_{k=\psi,\phi} {\langle k| \hat{\rho}^S_{i,\delta_T}|k \rangle} \frac{1}{2i}\left(\frac{{\langle k|\hat{A} \hat{\rho}^S_{i,\delta_T}|k \rangle}}{{\langle k| \hat{\rho}^S_{i,\delta_T}|k \rangle}}- \frac{{\langle k| \hat{\rho}^S_{i,\delta_T}\hat{A}|k \rangle}}{{\langle k| \hat{\rho}^S_{i,\delta_T}|k \rangle}}\right)=0.\label{eq: re1}
\end{align}
By using a conservation of the probability $\sum_{k=\psi,\phi} {\langle k| \hat{\rho}^S_{i,\delta_T}|k \rangle}=1$ and Eq. (\ref{eq: re1}), we also get following relations:
\begin{align}
&{\langle \phi|\hat{\rho}^S_{i,\delta_T}|\phi\rangle}=\frac{{\rm Im} A^W_{\psi}}{{\rm Im}A^W_{\psi} -{\rm Im}A^W_{\phi}},\label{eq: re2}
\\
&{\langle \psi|\hat{\rho}^S_{i,\delta_T}|\psi\rangle}=\frac{{\rm Im} A^W_{\phi}}{{\rm Im}A^W_{\phi} -{\rm Im}A^W_{\psi}}.\label{eq: re3}
\end{align}
Thanks to Eq. (\ref{eq: re1}), Eq. (\ref{eq: re2}) and Eq. (\ref{eq: re3}), the entropy difference is expressed by the weak values.
%

%\begin{widetext}
\section{Calculation of difference in von Neumann entropy}
In this section, we show detail calculations of the difference in von Neumann entropy between the initial and final state.
%value
First, we evaluate the von Neumann entropy at the initial time $t_i$ as
\begin{align}
S(\hat{\rho}_i)&=-{\rm Tr} \left(\hat{\rho}_i \ln \hat{\rho}_i \right)
\\
&=-{\rm Tr}_M \sum_{k=\psi,\phi} {\langle k|\hat{\rho}_i \ln \hat{\rho}_i}{|k\rangle}\notag
\\
&=S(\hat{\rho}^S_i) +S(\hat{\rho}^M_i),\label{eq: enini}
\end{align}
where ${\rm Tr}_P$ is the trace defined on the Hilbert space $\mathcal{H}_M$ and
\begin{align}
S(\hat{\rho}^M_i)&=\ln \left[e^{1/2} L \sqrt{\frac{1}{2\pi d^2}}\right].\label{eq: enini2}
\end{align}

Similarly, the von Neumann entropy at the final time $t_f$ is given as
\begin{align}
S(\hat{\rho}_4)&=-{\rm Tr} (\hat{\rho}_4 \ln \hat{\rho}_4)
\\
&=-{\rm Tr}_M \sum_{k=\psi,\phi} {\langle k|}\hat{\rho}_4 \ln \hat{\rho}_4 {|k\rangle}\notag
\\
&=-{\rm Tr}_M \left[\sum_{k=\psi,\phi} p_k \hat{\rho}^{(k)}_4 \left(\ln p_k +\ln \hat{\rho}^{(k)}_4 \right) \right]\notag
\\
&=H(\{p_k\}) +\sum_{k=\psi,\phi} p_k S(\hat{\rho}^{(k)}_4),\label{eq: enf}
\end{align}
where, for convenience, we defined following states:
\begin{align}
\hat{\rho}^{(k)}_4\equiv\frac{{\rm exp}\left(-d_k^2 \hat{p}^2_{L_k}/2 \right)}{Z_{d_k,L_k}},~~Z_{d_k,L_k}\equiv {\rm Tr}_M \left({\rm exp}\left(-d_k^2 \hat{p}^2_{L_k}/2\right) \right).
\end{align} 
The Shannon information content $H(\{p_k\}) $ is evaluated as
\begin{align}
H(\{p_k\})&=-\sum_{k=\psi,\phi}p_k \ln p_k
\\
&=-\sum_{k=\psi,\phi}{{\langle k| \hat{\rho}^S_{i}|k\rangle}}\left(1+g T{\rm Im}[A^W_k] \right)  \ln \left[{{\langle k| \hat{\rho}^S_{i}|k\rangle}}\left(1+gT {\rm Im}[A^W_k] \right)\right] +\mathcal{O}(g^2)\notag
\\
&=-\sum_{k=\psi,\phi} {{\langle k| \hat{\rho}^S_{i}|k\rangle}} \ln {{\langle k| \hat{\rho}^S_{i}|k\rangle}}-gT\sum_{k=\psi,\phi}{\rm Im}A^W_k\cdot {\langle k| \hat{\rho}^S_{i}|k\rangle} \ln {\langle k| \hat{\rho}^S_{i}|k\rangle}
-gT \sum_{k=\psi,\phi}{{\langle k| \hat{\rho}^S_{i}|k\rangle}}\cdot {\rm Im}A^W_k +\mathcal{O}(g^2)\label{eq: en1}
\\
&=H(\{{\langle k| \hat{\rho}^S_{i}|k\rangle}\})-gT\frac{{\rm Im}A^W_{\psi}{\rm Im} A^W_{\phi}}{{\rm Im}A^W_{\phi}-{\rm Im}A^W_{\psi}} \ln \left[ - \frac{{\rm Im}A^W_{\phi}}{{\rm Im}A^W_{\psi}}\right]+\mathcal{O}(g^2).\label{eq: en2}
\end{align}
Here, the second term in Eq.(\ref{eq: en2}) was given by using Eq. (\ref{eq: re2}) and Eq. (\ref{eq: re3}) and the third term in Eq. (\ref{eq: en1}) do not contribute because of Eq. (\ref{eq: re1}).

Besides, the von Neumann entropies $S(\hat{\rho}^{(k)}_4)$ are given as
\begin{align}
S(\hat{\rho}^{(k)}_4)=\ln \left[e^{1/2} L_k \sqrt{\frac{1}{2\pi d_k^2}}\right],~~~{\rm for}~k=\psi,\phi.\label{eq: en3}
\end{align}
%
%Since, in the weak measurement with the post-selection, mechanical operations corresponding to a change of the volume of gas are not included ordinary, we assumed that the size of the one-dimensional space is not changed before and after the post-selection.

By using Eq.(\ref{eq: enini}), Eq.(\ref{eq: enini2}),  Eq.(\ref{eq: enf}), Eq.(\ref{eq: en2}) and Eq.(\ref{eq: en3}), the difference in von Neumann entropy between the initial and final state is calculated as
\begin{align}
S(\hat{\rho}_4)-S(\hat{\rho}_1)&=p_{\psi}\ln \left[\frac{L_{\psi}({\rm Im}~A^W_{\phi}-{\rm Im}~A^W_{\psi}) }{L \cdot{\rm Im}~A^W_{\phi}} \right]+p_{\phi}\ln \left[\frac{L_{\phi}({\rm Im}~A^W_{\psi}-{\rm Im}~A^W_{\phi}) }{L \cdot{\rm Im}~A^W_{\psi}} \right]  -S(\hat{\rho}^S_i)+\mathcal{O}(g^2)
\notag \\[2mm]
&=-S(\hat{\rho}^S_i)+\mathcal{O}(g^2).
\end{align}

This is actuary equals to the QC mutual information content of our system.
%
%For the maximized weak values,  

%\section{QC-mutual information content}
%\label{sec:ApC}
%
%In this section, we show detail calculations of the QC-mutual information content.
%
%First, we show that the lower bound of difference in von Neumann entropy is characterized by the QC-mutual information.
%
%The difference in von Neumann entropy is expressed as
%
%\begin{align}
%S(\hat{\rho}_f) -S(\hat{\rho}_i)&=S\left(\sum_k p_k \hat{\rho}^{(k)}(L) \right)-S(\hat{\rho}_i)
%\\
%&=-\sum_k p_k \ln p_k +\sum_k p_k S\left(\hat{\rho}^{(k)}(L)\right)-S(\hat{\rho}_i)\label{eq: li2}
%\\
%&\geq \sum_k p_k S\left(\hat{\rho}^{(k)}(L)\right)-S(\hat{\rho}_i)\equiv -I (\hat{\rho}_i,\{p_{\psi},p_{\phi} \}).
%\end{align}  
%
%The QC-mutual information in this system is given as
%
%\begin{align}
%I (\hat{\rho}_i,\{p_{\psi},p_{\phi} \})= S(\hat{\rho}_i^E)+\mathcal{O}(g^2).
%\end{align}

\end{widetext}


\begin{thebibliography}{99}

%\cite{Glashow:1970gm}

\bibitem{PhysRev.134.B1410}
{Y.~Aharonov, P.~G.~Bergmann, and J.~L.~Lebowitz},
{``Time Symmetry in the Quantum Process of Measurement'},
\href{https://link.aps.org/doi/10.1103/PhysRev.134.B1410}{Phys.\ Rev.\ B {\bf 134}, 1410 (1964)}


\bibitem{Aharonov:1988ho}
Y.~Aharonov, D.~Z.~Albert, and L.~Vaidman,
  ``How the result of a measurement of a component of the spin of a spin-1/2 particle can turn out to be 100,''
  \href{https://link.aps.org/doi/10.1103/PhysRevLett.60.1351}{Phys.\ Rev.\ Lett. {\bf 60}, 1351 (1988).}

\bibitem{Ritchie:1991}
  N.~W.~M.~ Ritchie, J.~G.~Story and Randall~G.~Hulet,
 ``Realization of a measurement of a ``weak value''"
  \href{https://link.aps.org/doi/10.1103/PhysRevLett.66.1107}{Phys.\ Rev.\ Lett. {\bf66}, 1107(1991).}
  
\bibitem{Pryde:2005}
  G.~J.~Pryde, J.~L.~O'Brien, A.~G.~White, T.~C.~Ralph, and H.~M.~Wiseman,
 ``Measurement of Quantum Weak Values of Photon Polarization"
  \href{https://link.aps.org/doi/10.1103/PhysRevLett.94.220405}{Phys.\ Rev.\ Lett. {\bf94}, 220405(2005).}
    
\bibitem{Hosten787}    
 O.~Hosten and P.~Kwiat,   
 { ``Observation of the Spin Hall Effect of Light via Weak Measurements"},
\href{https://science.sciencemag.org/content/319/5864/787}{Sience {\bf 319}, 787--790 (2008).}

\bibitem{PhysRevLett.102.173601}
P.~B.~Dixon, D.~J.~Starling, A.~N.~Jordan, J.~C.~Howell,
{``Ultrasensitive Beam Deflection Measurement via Interferometric Weak Value Amplification"},
\href{https://link.aps.org/doi/10.1103/PhysRevLett.102.173601}{Phys.\ Rev.\ Lett. {\bf102}, 173601(2009).}
  author = {Dixon, P. Ben and Starling, David J. and Jordan, Andrew N. and Howell, John C.}
  
  \bibitem{Sagawa:2008se} 
T.~Sagawa and M.~Ueda,
  ``Second Law of Thermodynamics with Discrete Quantum Feedback Control,''
  \href{https://link.aps.org/doi/10.1103/PhysRevLett.100.080403}{Phys.\ Rev.\ Lett. {\bf 100}, 8 (2008).}
  %%CITATION = doi:10.1103/PhysRevD.2.1285;%%

\bibitem{Sagawa:2010ge}
T.~Sagawa and M.~Ueda,
  ``Generalized Jarzynski Equality under Nonequilibrium Feedback Control,''
  \href{https://link.aps.org/doi/10.1103/PhysRevLett.104.090602}{Phys.\ Rev.\ Lett. {\bf 104}, 090602 (2010).}
    

\bibitem{Seifert:2012} 
U.~Seifert,
  ``Stochastic thermodynamics, fluctuation theorems and molecular machines,''
  \href{https://doi.org/10.1088}{Rep.\ Prog.\ Phys. {\bf 75}, 126001 (2012)}
  
 \bibitem{Evans:1993} 
D.~J.~Evans, E.~G.~D.~Cohen and G.~P.~Morriss,
  ``Probability of second law violations in shearing steady states,''
  \href{https://link.aps.org/doi/10.1103/PhysRevLett.71.2401}{Phys.\ Rev.\ Lett. {\bf 71}, 2401 (1993)} 
  
 \bibitem{Gallavotti:1995} 
G.~Gallavotti and E.~G.~D.~Cohen,
  ``Dynamical Ensembles in Nonequilibrium Statistical Mechanics,''
  \href{https://link.aps.org/doi/10.1103/PhysRevLett.74.2694}{Phys.\ Rev.\ Lett. {\bf 74}, 2694 (1995)} 

 \bibitem{Jarzynski:1997} 
C.~Jarzynski,
  ``Nonequilibrium Equality for Free Energy Differences,''
  \href{https://link.aps.org/doi/10.1103/PhysRevLett.78.2690}{Phys.\ Rev.\ Lett. {\bf 78}, 2690 (1997)} 

 \bibitem{Crooks:1999} 
G.~E.~Crooks,
  ``Entropy production fluctuation theorem and the nonequilibrium work relation for free energy differences,''
  \href{https://link.aps.org/doi/10.1103/PhysRevE.60.2721}{Phys.\ Rev.\ E {\bf 60}, 2721 (1999)} 

 \bibitem{Jarzynski:2000} 
C.~Jarzynski,
  ``Hamiltonian Derivation of a Detailed Fluctuation Theorem,''
 {J.\ Stat.\ Phys. {\bf 98}, 77 (2000)} 

 \bibitem{Hatano:2001} 
T.~Hatano and S.-I.~Sasa,
  ``Steady-State Thermodynamics of Langevin Systems,''
  \href{https://link.aps.org/doi/10.1103/PhysRevLett.86.3463}{Phys.\ Rev.\ Lett. {\bf 86}, 3463 (2001)} 

 \bibitem{Seifert:2005} 
U.~Seifert,
  ``Entropy Production along a Stochastic Trajectory and an Integral Fluctuation Theorem,''
  \href{https://link.aps.org/doi/10.1103/PhysRevLett.95.040602}{Phys.\ Rev.\ Lett. {\bf 95}, 040602 (2005)} 

\bibitem{Toyabe:2010ex}
S.~Toyabe, T.~Sagawa, M.~Ueda, E.~Muneyuki and M.~Sano,
  ``Experimental demonstration of information-to-energy conversion and validation of the generalized Jarzynski equality,''
  \href{https://doi.org/10.1038/nphys1821}{Nature {\bf 6}, 988 (2010).}

\bibitem{Maxwell:1871} 
J.~C.~Maxwell,
  ``Theory of Heat,''
{(Appleton, London, 1871).}

 \bibitem{Szilard:1929} 
L.~Szilard,
{Z.\ Phys. {\bf 53}, 840 (1929)}

 \bibitem{Landauer:1961} 
R.~Landauer,
  ``Irreversibility and Heat Generation in the Computing Process,''
{IBM\ J.\ Res.\ Dev. {\bf 5}, 183 (1961)} 

 \bibitem{Bennett:1982} 
C.~H.~Bennett,
  ``The thermodynamics of computation---a review,''
{Int.\ J.\ Theor.\ Phys. {\bf 21}, 905 (1982)} 

 \bibitem{Piechocinska:2000} 
B.~Piechocinska,
  ``Information erasure,''
  \href{https://link.aps.org/doi/10.1103/PhysRevA.61.062314}{Phys.\ Rev.\ A {\bf 61}, 062314 (2000)} 
 
  \bibitem{Lloyd:1997} 
S.~Lloyd,
  ``Quantum-mechanical Maxwell's demon,''
  \href{https://link.aps.org/doi/10.1103/PhysRevA.56.3374}{Phys.\ Rev.\ A {\bf 56}, 3374 (1997)} 

  \bibitem{Nielsen:1998} 
M.~A.~Nielsen, C.~M.~Caves, B.~Schumacher and H.~Barnum,
  ``Information-theoretic approach to quantum error correction and reversible measurement,''
  \href{https://doi.org/10.1098/rspa.1998.0160}{Proc.\ R.\ Soc.\ A {\bf 454}, 277 (1998)} 

  \bibitem{Maruyama:2005} 
  K.~Maruyama, F.~Morikoshi and V.~Vedral,
  ``Thermodynamical detection of entanglement by Maxwell's demons,''
  \href{https://link.aps.org/doi/10.1103/PhysRevA.71.012108}{Phys.\ Rev.\ A {\bf 71}, 012108 (2005)} 

  \bibitem{Maruyama:2009} 
  K.~Maruyama, F.~Nori and V.~Vedral,
  ``The physics of Maxwell's demon and information,''
  \href{https://link.aps.org/doi/10.1103/RevModPhys.81.1}{Rev.\ Mod.\ Phys.\ {\bf 81}, 1 (2009)} 

\bibitem{10.1093/ptep/ptz019}
K.~Kawana and D.~Ueda,
  ``Amplification of gravitational motion via quantum weak measurement,''
  \href{https://doi.org/10.1093/ptep/ptz019}{PTEP {\bf 2019}, 041A01 (2019)}

\bibitem{Rosi:2017qu}
G.~Rosi, G.~D’Amico, L.~Cacciapuoti, F.~Sorrentino, M.~Prevedelli, M.~Zych, C.~Brukner and G.~M.~Tino, 
  ``Quantum test of the equivalence principle for atoms in coherent superposition of internal energy states,''
  \href{https://doi.org/10.1038/ncomms15529}{Nature {\bf 8}, 15529 (2017).}


  
\end{thebibliography}
\end{document}